\documentclass[12pt]{iopart}

\usepackage{graphicx}
\usepackage{mathrsfs}
\usepackage{amssymb}
\usepackage{bm}
\usepackage{amsbsy}
\usepackage{amssymb}
\usepackage{epstopdf}
\usepackage{mathrsfs}
\usepackage{multirow}
\usepackage{xcolor}

\begin{document}
	
	\title{Fate of localization features in a one-dimensional non-Hermitian flat-band lattice with quasiperiodic modulations}
	
	\author{Hui Liu}
	\address{Institute of Theoretical Physics and State Key Laboratory of Quantum Optics and Quantum Optics Devices, Shanxi University, Taiyuan 030006, P.R.China}
	\author{Zhanpeng Lu}
	\address{Institute of Theoretical Physics and State Key Laboratory of Quantum Optics and Quantum Optics Devices, Shanxi University, Taiyuan 030006, P.R.China}
	\author{Xu Xia}
	\address{Academy of Mathematics and System Sciences, Chinese Academy of Sciences, Beijing 100190, China}
	\author{Zhihao Xu*}
	\address{Institute of Theoretical Physics and State Key Laboratory of Quantum Optics and Quantum Optics Devices, Shanxi University, Taiyuan 030006, P.R.China}
	\address{Collaborative Innovation Center of Extreme Optics, Shanxi University, Taiyuan 030006, China}
	\address{*Author to whom any correspondence should be addressed.}
	\ead{xuzhihao@sxu.edu.cn}

	\begin{abstract}
		We investigate the influence of quasiperiodic modulations on one-dimensional non-Hermitian diamond lattices with an artificial magnetic flux $\theta$ that possess flat bands. Our study shows that the symmetry of these modulations and the magnetic flux $\theta$ play a pivotal role in shaping the localization properties of the system. When $\theta=0$, the non-Hermitian lattice exhibits a single flat band in the crystalline case, and symmetric as well as antisymmetric modulations can induce accurate mobility edges. In contrast, when $\theta=\pi$, the clean diamond lattice manifests three dispersionless bands referred to as an "all-band-flat" (ABF) structure, irrespective of the non-Hermitian parameter. The ABF structure restricts the transition from delocalized to localized states, as all states remain localized for any finite symmetric modulation. Our numerical calculations further unveil that the ABF system subjected to antisymmetric modulations exhibits multifractal-to-localized edges. Multifractal states are predominantly concentrated in the internal region of the spectrum. Additionally, we explore the case where $\theta$ lies within the range of $(0, \pi)$, revealing a diverse array of complex localization features. Finally, we propose a classical electrical circuit scheme to realize the non-Hermitian flat-band chain with quasiperiodic modulations.
	\end{abstract}
	
	
	
	\maketitle
	\section{Introduction}
	Anderson localization is a fundamental quantum phenomenon in which quantum waves become localized due to disorder \cite{Anderson1958} in their environment. In a three-dimensional (3D) scenario with uncorrelated disorder, the system exhibits an energy-dependent transition from extended to localized eigenstates. This critical energy level, denoted as $E_c$, is known as the mobility edge. The mobility edge plays a crucial role in shaping the properties and behavior of the system, including its conductivity \cite{YWang2022} and thermoelectric response \cite{Chiaracane2020}. In contrast to traditional Anderson models with uncorrelated disorder, where even a minuscule amount of disorder leads to complete localization in 1D and 2D cases, the 1D Aubry-Andr\'{e} (AA) model with a quasiperiodic modulation demonstrates a metal-insulator transition at a finite value of the onsite potential's amplitude \cite{Aubry1980} without mobility edges. Recently, various generalized AA models have been devised \cite{Biddle2010,Ganeshan2015,Wang2020,Xu2022}, which can display exact mobility edges. Importantly, quasiperiodic systems can give rise to a third category of states known as multifractal states \cite{Dai2023}, which exhibit both extended and non-ergodic properties. Consequently, in addition to the mobility edge, a novel type of mobility edge between multifractal and localized states has been proposed \cite{Liu2022,Zhou2023}. This concept holds significant importance in developing models with multifractal states.
	
	On the other hand, localization can also be attained in the absence of disorder, particularly in certain translation-invariant systems with energy bands that lack dispersion, denoted as flat bands  \cite{Rhim2021,Talkington2022,Morfonios2021,Kuno2020,Zhang2021,Ramezani2017,Sathe2309,Green2010}. These flat bands are characterized by having energy levels independent of the momentum, $E(k)=E$, resulting in a large-scale degeneracy at the energy $E$. This extensive degeneracy leads to the presence of compact localized states (CLSs) within the flat bands \cite{Green2010}, where the eigenstates are confined to a finite number of sites \cite{Sutherland1986,Aoki1996}.  CLSs have been observed in specially engineered lattices, including cross-stitch \cite{Gne2018,Toy2018}, diamond \cite{Vidal2000,Muk2018}, kagome \cite{Chern2014}, dice \cite{Kolo2018}, and pyrochlore lattices \cite{Mizoguchi2019}. 
	
	Systems featuring flat bands are of significant interest due to their potential to exhibit exotic and emergent phenomena. These systems often display strong correlations, giving rise to unconventional phases of matter, such as high-temperature superconductivity \cite{Tian2023}, unconventional magnetism \cite{Tasaki1992}, or topologically non-trivial states \cite{Peo2015}. Recent theoretical studies have explored the introduction of a disordered potential to break the macroscopic degeneracy in flat-band systems \cite{Longhi2021,Wang2022,Li2022,WZhang2023}. By incorporating quasiperiodic modulations into certain flat-band geometries \cite{Bodyfelt2014,Daneli2015}, precise engineering and fine-tuning of mobility edges become possible. Notably, when a small amount of quasiperiodic AA disorder is introduced to a compactly localized ABF diamond chain, the resulting eigenstates exhibit multifractality \cite{Ahmed2022,Lee2023,Lee}, and an exact transition from multifractal to localized states is observed.
	
	In recent years, non-Hermitian systems have garnered considerable attention in both experimental and theoretical domains. Among these, non-reciprocal systems have emerged as a particularly intriguing area of study, appearing in various forms across physics. These systems exhibit remarkable properties that have no counterparts in reciprocal systems, including the non-Hermitian skin effect \cite{Yao2018,SYao2018,Gong2018,Xu2021} and novel topological features in the complex spectrum \cite{Geng2023,LiJ2024}. The interplay between non-Hermiticity and disorder has ignited a fresh perspective on localization characteristics. For instance, in the Hatano-Nelson model \cite{Gong2018,Hatano1996,Hatano1997,Hatano1998}, when nonreciprocal hopping is combined with uncorrelated disorder, a finite transition from extended to localized states is observed. In generalized non-Hermitian AA models, a simultaneous occurrence of real-complex transition, topological phase transition, and localization transition has been identified \cite{Xu2021}. Moreover, exactly solvable non-Hermitian quasiperiodic models have been proposed for both 1D and 2D systems \cite{Xu2022,Longhis2021}. Non-Hermitian localization and delocalization phenomena in two-dimensional photonic quasicrystals within atomic and atomic-like ensembles have also been studied both theoretically and experimentally \cite{Feng2023,Zhang2024}. These intriguing developments in non-Hermitian disordered systems have led to the consideration of non-Hermitian effects in flat-band models with quasiperiodic modulations. Such modulated non-Hermitian flat-band systems hold promise for realization in various experimental platforms, including electrical circuits \cite{Helbig2020,Li2024,LiuS2021,ZouD2021,ZhangX2021}, acoustic \cite{ZhangX2021,ZhangL2021,Gao2022} and photonic lattices \cite{Weidemann2020,SongY2020}, single-photon quantum walks \cite{LinQ2022,XiaoL2020,XiaoL2021}, and cold atoms \cite{Li2022}.
	
	In this paper, we systematically investigate the impact of quasiperiodic modulations on a diamond lattice featuring flat bands with nonreciprocal hoppings. Our findings reveal that the symmetry of the external modulations and the synthetic magnetic flux parameter, $\theta$, play a pivotal role in determining the localization properties of the system. We provide a comprehensive analysis of the localization characteristics of the non-Hermitian flat-band chain under varying symmetries of quasiperiodic modulation and synthetic magnetic flux. Additionally, we propose an experimental scheme using electrical circuits to realize our non-Hermitian model.
	
	\section{Model and Hamiltonian}
	
	\begin{figure}[!htb]
		\begin{center}
			\includegraphics[scale=0.4]{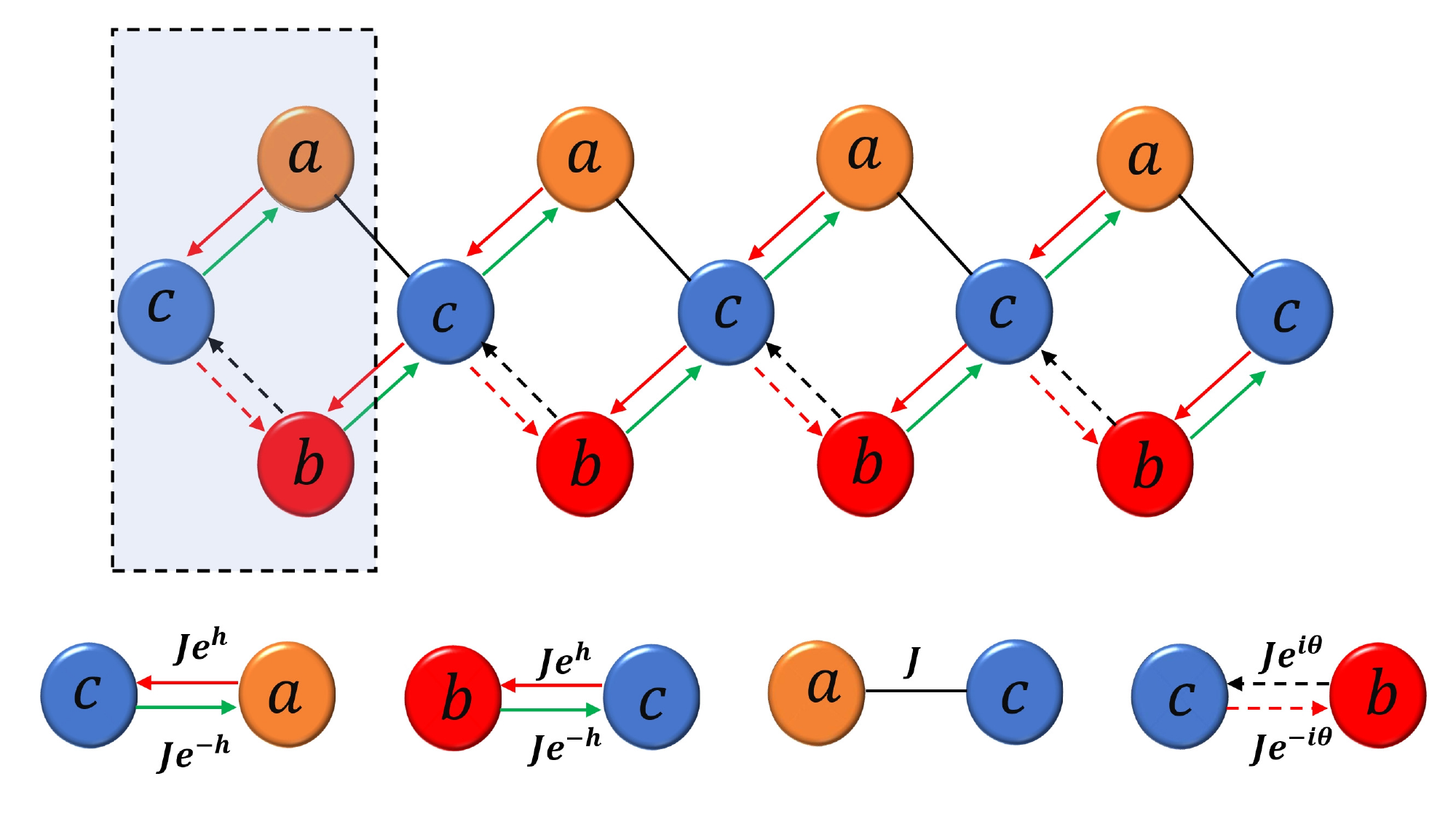}
		\end{center}
		\caption{(Color online) Schematic diagram of a non-Hermitian diamond lattice chain.}\label{Fig1}
	\end{figure}
	
	We consider a non-Hermitian diamond chain with quasiperiodic modulations. In the clean case as schematically illustrated in Fig.~\ref{Fig1}, there are three sublattices labeled by $a$, $b$, and $c$ in each unit cell. We introduce nonreciprocal couplings marked by solid-line arrows between sublattices $a$ and $c$ in the same unit cell, and sublattices $b$ and $c$ in adjacent unit cells. Moreover, a synthetic magnetic flux $\theta$ is applied in each closed diamond loop via Peierls' substitution of the coupling constant between sublattices $b$ and $c$ in the same unit cell. The phase acquired by the wave function of the particle as it runs a closed diamond loop containing a nonzero magnetic flux, which has been experimentally demonstrated in electronic systems \cite{Martinez2023}, photonics \cite{Kremer2020}, and cold atoms \cite{Li2022}. Consider the eigenvalue problem of a generalized tight-binding model
	\begin{equation}
		E\psi_n=-J\left[\hat{V}\psi_n+\hat{T}_1\psi_{n-1}+\hat{T}_2\psi_{n+1}\right] +\hat{\epsilon}_n\psi_n,
		\label{eq1}
	\end{equation}
	where
	\begin{eqnarray}
		\hat{V}=\left(\begin{array}{ccc}0&0&e^{-h}\\0&0&e^{-i\theta}\\e^h&e^{i\theta}&0\end{array}
		\right),\hat{T}_1=\left(\begin{array}{ccc}0&0&0\\0&0&0\\1&e^{-h}&0\\ \end{array}
		\right),
		\hat{T}_2=\left(\begin{array}{ccc}0&0&1\\0&0&e^h\\0&0&0\end{array}
		\right),
		\label{eq2}
	\end{eqnarray}
    represent the intracell couplings, the couplings with the left unit cell, and the couplings with the right unit cell, respectively. Here,  each component of the vector $\psi_n=(a_n,b_n,c_n)^T$ represents a site of a periodic lattice in the $n$-th unit cell, $J$ is the coupling amplitude between adjacent sites with $J=1$ being set as the unit of energy. $h$ is a virtual gauge potential leading to nonreciprocal hoppings and non-Hermitian phenomena in the system. $\theta\in[0,\pi]$ is the synthetic magnetic flux, and the unit cell modulation $\hat{\epsilon}_n$ of the Hamiltonian (\ref{eq1}) is given by the diagonal square matrix $\hat{\epsilon}_n=\mathrm{diag}(\epsilon_n^a,\epsilon_n^b,\epsilon_n^c)$. 
	
	\begin{figure}[!htb]
		\begin{center}
			\includegraphics[scale=0.6]{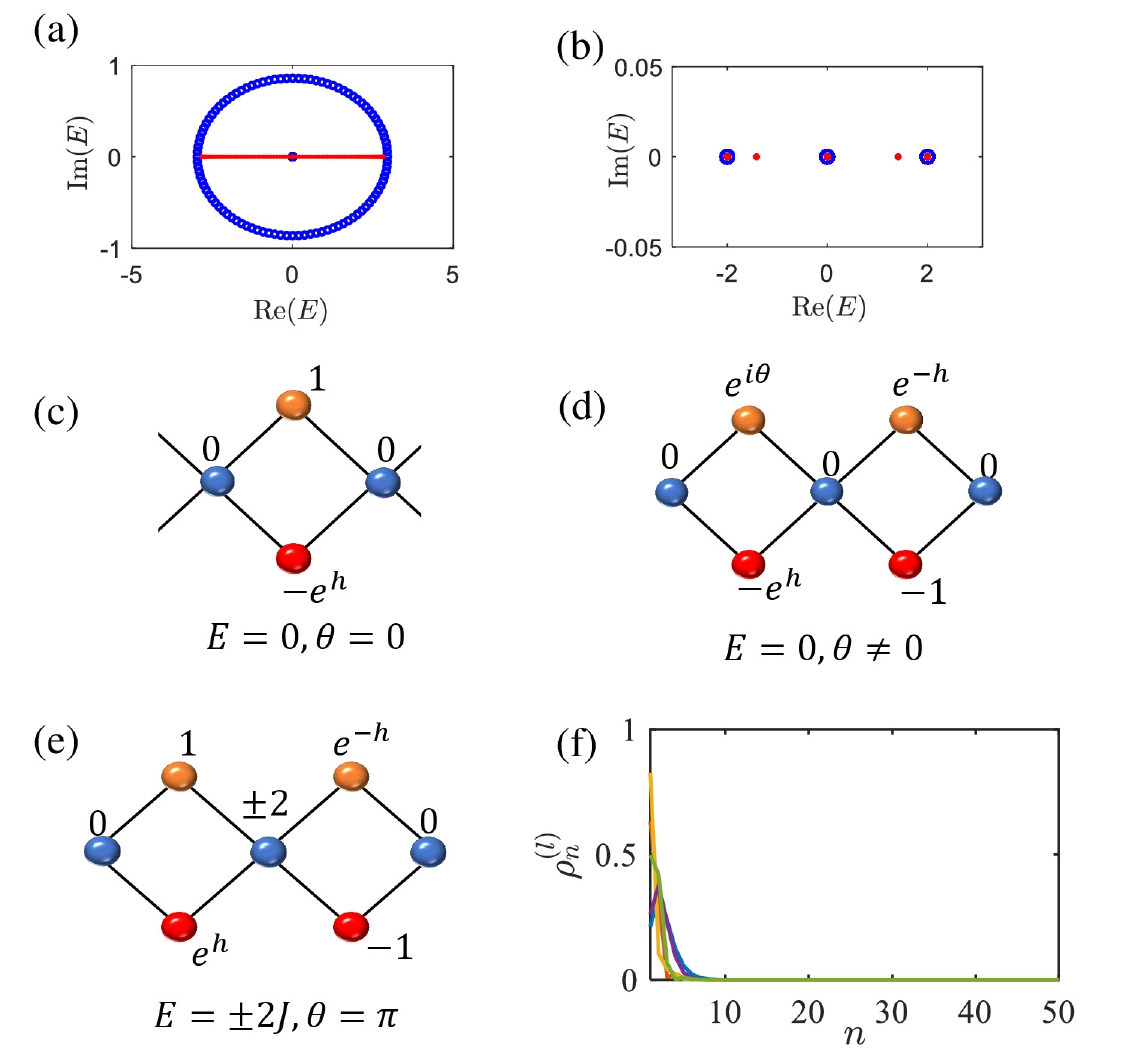}
		\end{center}
		\caption{(Color online) Crystalline Case: Complex energy spectra for (a) $\theta=0$ and (b) $\theta =\pi$ under different boundary conditions. Solid red dots and blue circles represent the theoretical results under PBC and OBC, respectively. (c)-(e) The CLS occupations of the non-Hermitian diamond lattice in different flat bands for different $\theta$. (f) Density distributions $\rho_n^{(l)}$ for six randomly selected eigenstates in dispersive bands with $\theta=0$ under OBCs.}\label{Fig2}
	\end{figure}
	In the crystalline case, where the on-site potential is set to be zero, the clean non-Hermitian model possesses three energy bands under periodic boundary conditions (PBCs) with the dispersion relations given by
	\begin{equation}
		E_0=0,\quad E_{\pm}(k)=\pm 2J\sqrt{1+\cos{\frac{\theta}{2}}\cos(\frac{\theta}{2}+k-ih)},
		\label{eq3}
	\end{equation}
	where $-\pi< k \leq\pi$ is the wave number. For $\theta \neq \pi$, there is only one flat band at energy $E_0$, and the other two energy bands are $k$-dependent. We plot the energy spectra of $\theta=0$ and $\pi$ under PBCs marked by blue circles and open boundary conditions (OBCs) marked by red points in Fig. \ref{Fig2}(a) and \ref{Fig2}(b), respectively. For $\theta=0$ under PBCs, all energies except $E_0$ are complex, whereas under OBCs, all energies become real. For $\theta=\pi$ under PBCs, the energy bands exhibit a remarkable dispersionless nature, showcasing energies of $E_0=0$ and $E_{\pm}=\pm 2J$. The energy spectrum remains real under OBCs for $\theta=\pi$. Like the Hermitian case, eigenmodes of our non-Hermitian model corresponding to those $k$-independent energies are CLSs whose amplitudes are nonzero only across a finite number of sites. Figure \ref{Fig2}(c) illustrates the fundamental CLSs for the $E_0=0$ state with $\theta=0$ in our cases, which occupies two sites and is localized in a single unit cell. When $\theta \ne 0$ and $E_0=0$, the fundamental CLSs occupy four sites, shown in Fig. \ref{Fig2}(d). The states of the additional flat bands at $E_{\pm}=\pm 2J$ for $\theta=\pi$ host five-site CLSs which excite one of the bottleneck sites seen in Fig. \ref{Fig2}(e). In Appendix A, we provide a detailed derivation of the CLSs and the corresponding density distribution under OBCs. Due to the existence of the nonreciprocal hoppings, the states in a dispersive band display the non-Hermitian skin effect under OBCs. To display the non-Hermitian skin effect, we calculate the $l$-th eigenstate's density distribution $\rho_n^{(l)} = \sum_{\beta\in \{ a,b,c\}}|\psi_{n,\beta}^{(l)}|^2$, where $\psi_{n,\beta}^{(l)}$ represents the normalized probability amplitude of the $\beta$ site in the $n$-th unit cell for the $l$-th eigenstate with the number of the unit cell being $N$, and the lattice size being $L=3N$. 
	In Fig. \ref{Fig2}(f), we show $\rho_n^{(l)}$ for six randomly selected eigenstates in dispersive bands with $h=0.6$, $N=50$, and $\theta=0$ under OBCs. According to Fig. \ref{Fig2}(f), one can see that different states of dispersive bands show the non-Hermitian skin effect.
	
	The effect of quasiperiodic AA modulations is considered in the present work. The onsite modulations $\{\epsilon_n^{\beta} \}$ for $\beta=\{a,b\}$ are defined as independent AA modulations $\epsilon_n^{\beta}=\lambda_{\beta} \cos(2\pi\alpha n+\phi_{\beta}),$ where the parameters $\lambda_{\beta}$ are positive real values controlling the quasiperiodic modulation amplitude, $\phi_{\beta}$ is the phase shift, and $\alpha$ is an irrational number which is set to be the golden ratio $\alpha=(\sqrt{5}-1)/2$. Without loss of generality, we set the $a$-leg phase to be zero ($\phi_a=0$). Moreover, the $a$- and $b$-leg modulation amplitudes are set to be equal to each other $\lambda_{\alpha}=\lambda_{\beta}=\lambda \ge 0$. The $c$-leg potential is a uniform modulation with the amplitude $\epsilon_n^c=K\in \mathbb{R}$.
	
	We utilize a local coordinate transformation to the unit cells, which rotates these lattices into a Fano defect form \cite{Bodyfelt2014,AE2010,Flach2014}. The rotation for our non-Hermitian system is defined by a real matrix $\hat{U}$:
	\begin{eqnarray}
		\left(\begin{array}{ccc}	p_n \\ f_n \\ c_n \\\end{array}
		\right)=\hat{U}
		\left(\begin{array}{ccc}a_n \\ b_n \\ c_n \\\end{array}
		\right) \quad
		\hat{U}=\frac{1}{\sqrt{2}}\left(\begin{array}{ccc}1&e^{-h}&0\\e^h&-1&0\\0&0&\sqrt{2}\end{array}\right),
		\label{eq4}
	\end{eqnarray}
	with $\varphi_n=(p_n,f_n,c_n)^T$ being the rotated tight-binding representation of wave function of the $n$-th unit cell. Additional details about $\hat{U}$ are provided in Appendix B. Lastly, such local coordinate transformation also rotates the onsite modulations. For the diamond chain, this gives
	\begin{equation}
		\hat{\tilde{\epsilon}}_n=\hat{U}\hat{\epsilon}_n\hat{U}^{-1}=\left(\begin{array}{ccc}\epsilon_n^{+}&\epsilon_n^{-}e^{-h}&0\\\epsilon_n^{-}e^h&\epsilon_n^{+}&0\\0&0&K\end{array}\right),
		\label{eq5}
	\end{equation}
	with $\epsilon_n^{\pm}=(\epsilon_n^a\pm\epsilon_n^b)/2$. From Eq.(\ref{eq5}), the remarkable correlations between the $a$- and $b$-leg modulations appear and will be an object of our studies in this work; namely
	\begin{eqnarray}
		Symmetric: \quad &\phi_b=0 & \Leftrightarrow \quad  \epsilon_n^{-}=0, \nonumber \\
		Antisymmetric: \quad  &\phi_b=\pi & \Leftrightarrow \quad  \epsilon_n^{+}=0.
		\label{eq6}
	\end{eqnarray}
	Since we have set the $a$-leg phase to be zeroed, according to Eq.(\ref{eq6}), the correlations can be obtained from the $b$-leg phase, {\it i.e.}, $\phi_b=0$ ($\phi_b=\pi$) for the symmetric (antisymmetric) case.
	According to Eq.(\ref{eq4}), the diamond lattice's Eq.(\ref{eq1}) become
	\begin{eqnarray}
		&Ep_n=\epsilon_n^+ p_n+\epsilon_n^- e^{-h}f_n-\frac{1}{\sqrt{2}}(1+e^{-i\theta})e^{-h}c_n-\sqrt{2}c_{n+1},\nonumber \\
		&Ef_n=\epsilon_n^+ f_n+\epsilon_n^- e^h p_n-\frac{1}{\sqrt{2}}(1-e^{-i\theta})c_n,\nonumber \\
		&(E-K)c_n=-\frac{1}{\sqrt{2}}\left[(1+e^{i\theta})e^h p_n+(1-e^{i\theta})f_{n}\right]-\sqrt{2}p_{n-1}.
		\label{eq7}	
	\end{eqnarray}
	Based on Eq.(\ref{eq7}), we discuss different choices of $\theta$ and correlations between the $a$- and $b$-leg modulations on the system's localization transitions. 
	
	To explore the localization properties of the eigenstates, one can calculate the $l$-th eigenstate's fractal dimension, which is defined as $D_2^{(l)}=-\ln I_2^{(l)}/\ln (L)$ with the inverse participation ratio (IPR) being $I_2^{(l)}=\sum_{n=1}^{N}\sum_{\beta\in \{ a,b,c\}}\left|\psi_{n,\beta}^{(l)}\right|^{4}$. For a localized state, $I_2^{(l)}=O(1)$ in the thermodynamic limit and the corresponding $D_2^{(l)} \to 0$, while for an extended state, $I_2^{(l)}$ tends to zero in the large system size limit and the corresponding $D_2^{(l)} \to 1$. For a multifractal wave function, $D_2^{(l)} \in (0,1)$ and the value of $I_2^{(l)}$ approaches zero in the $L\to\infty$ limit. 
	
	To further verify the existence of the multifractal region in our system, we apply the mean inverse participation ratio of a given region $\mathrm{MIPR}(\sigma_{\tilde{\beta}})=\frac{1}{\mathcal{N}_{\sigma_{\tilde{\beta}}}} \sum_{l\in \{\sigma_{\tilde{\beta}}\}} I_2^{(l)}$. Where $\sigma_{\tilde{E}}$, $\sigma_{\tilde{M}}$, and $\sigma_{\tilde{L}}$ represent the spectra localized in the extended, multifractal, and localized regions, respectively, $\tilde{\beta}\in\{\tilde{E},\tilde{M},\tilde{L}\}$, and $\mathcal{N}_{\sigma_{\tilde{\beta}}}$ is the total amount of eigenvalues belonging to $\sigma_{\tilde{\beta}}$. For a finite size system, we use the function $\mathrm{MIPR}(\sigma_{\tilde{\beta}})=\tilde{a} \times L^{-\tilde{b}}+\tilde{c}$ for fitting and obtain the fitting parameters $\tilde{a}$, $\tilde{b}$, and $\tilde{c}$. For a perfectly localized region, $\mathrm{MIPR}(\sigma_{\tilde{L}})$ maintains a constant that hardly changes with $L$. For a perfectly extended region, $\mathrm{MIPR}(\sigma_{\tilde{E}})$ varies linearly with $1/N$ and $\tilde{c}\to 0$. When we consider the multifractal region, we can find that the fitting parameter $\tilde{b}\in (0,1)$ with finite $\tilde{a}$, and the fitting parameter $\tilde{c}$ approaches zero.
	
	In addition, many extensions to the AA model have recently been applied to the non-Hermitian systems, where one has systematically examined the relationship between the real-complex transition in energy and the delocalization-localization transition \cite{Tang2021}. In a class of AA models with nonreciprocal hoppings under PBCs, one discovered that delocalized states correspond to the complex and localized states to the real energies. Applying such properties, one can also detect the localization transitions of this class of non-Hermitian AA systems with nonreciprocal hoppings.
	
	In the following paper, we utilize the exact diagonalization method to do our numerical calculations. We set $h=0.6$ and $K=0.7$ as a concrete example, and the PBCs are considered.
	
	\section{ Localization features}
	We first consider the $\theta=0$ cases for Eq.(\ref{eq7}). It can be reduced into a tight-binding form by expressing the $f_n$ and $c_n$ variables through the $p_n$ ones, which contains the $p_n$ variables only:
	\begin{equation}
		\left[\frac{E(E-K)}{2}-2\right] p_n =(e^{-h}p_{n-1}+e^{h}p_{n+1})+\tilde{\epsilon}_n^{(1)} p_n, \label{eq8}
	\end{equation}
	where the effective on-site potential
	\begin{equation}
		\tilde{\epsilon}_n^{(1)} =\frac{E-K}{2} \left[\epsilon_n^{+} + \frac{(\epsilon_n^{-})^2}{E-\epsilon_n^{+}}\right]\label{eq9}
	\end{equation}
	is a function of the two on-site energies of the diamond lattice $\epsilon_n^{a,b}$ and depends on the on-site energy $K$ of the $c$-chain. Notice that regardless of the other system parameters, at the energy $E=K$ for $\theta=0$, we have $p_n=f_n=0$ and $c_n=e^{h}c_{n+1}$, where the state's amplitudes reside on the $c$ sites. Hence, an extended state exists under PBCs and a non-Hermitian skin state can be observed under OBCs at the energy $E=K$, independent of the modulation strength $\lambda$.
	
\begin{figure}[!htb]
		\begin{center}
			\includegraphics[scale=1]{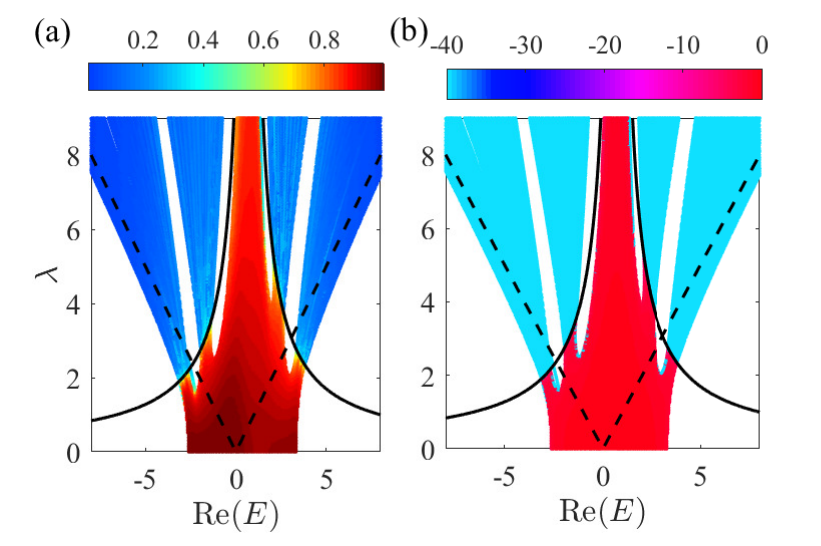}
		\end{center}
		\caption{(Color online) Symmetric Case with $\theta=0$: (a) The  real part of the spectrum $\sigma_{p,c}$ as a function of $\lambda$, where the color denotes the value of $D_2^{(l)}$. (b) $\ln{|\mathrm{Im}(E)|}$ as a function of $\lambda$ and $\mathrm{Re}(E)$, where the color denotes the value of $\ln{|\mathrm{Im}(E)|}$. The black solid lines represent the mobility edges given by Eq. (\ref{eq10}). The spectrum $\sigma_{f}$ is omitted, but its boundaries are indicated by black dashed lines. Here, $N=200$.}\label{Fig3}
	\end{figure}

\begin{figure}[!htb]
		\begin{center}
			\includegraphics[scale=1]{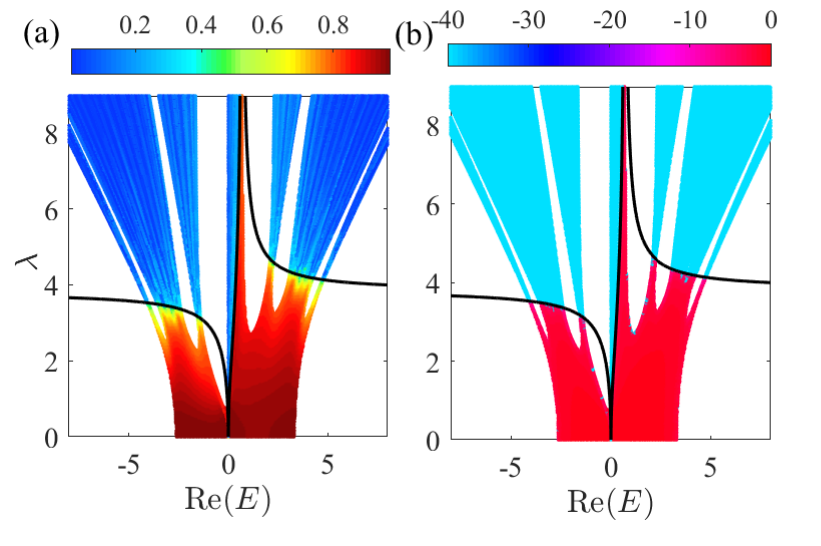}
		\end{center}
		\caption{(Color online) Antisymmetric Case with $\theta=0$: (a) The real part of the spectrum as a function of $\lambda$, where the color denotes the value of $D_2^{(l)}$. (b) $\ln{|\mathrm{Im}(E)|}$ as a function of $\lambda$ and $\mathrm{Re}(E)$, where the color denotes the value of $\ln{|\mathrm{Im}(E)|}$. The black solid lines represent the mobility edges given by Eq. (\ref{eq11}). Here, $N=200$.}\label{Fig4}
	\end{figure}

	In the $\theta=0$ case, the symmetric modulations $\epsilon_n^-=0$ can be obtained by setting $\phi_b=0$. The $f_n$ variables decouple from both the $p_n$ and $c_n$ variables, producing two independent spectra $\sigma_f$ and $\sigma_{p,c}$. The $\sigma_f$ keeps its compact feature with the energies given by $E=\epsilon_n^{+}$. Hence, all the states belonging to the spectrum $\sigma_f$ are localized. In Fig.\ref{Fig3}, we show the spectrum of the system with $\theta=0$ and $\phi_b=0$ as a function of $\lambda$. Due to the independence of the spectra $\sigma_f$ and $\sigma_{p,c}$, we indicate the boundaries of the spectrum $\sigma_f=\{ \epsilon_n^{+} \}$ by dashed lines in Fig.\ref{Fig3}, which is equidistributed within the interval $[-\lambda,\lambda]$. To obtain the localization property of $\sigma_{p,c}$, we can obtain the tight-binding form only containing the $p_n$ variables by choosing $\epsilon_n^-=0$ in Eq. (\ref{eq8}). The dispersive states $p_n$ are described by a non-Hermitian AA chain with the nonreciprocal hopping term. Referring to the discussion of the localization transition of the non-Hermitian AA model \cite{Jiang2019}, we can obtain the mobility edges
	\begin{equation}
		\lambda_c=\left |\frac{4M}{E_c-K}\right |, \label{eq10}
	\end{equation} 
	with $M=\max \{e^h,e^{-h}\}$. Figures \ref{Fig3}(a) and \ref{Fig3}(b) show the fractal dimension $D_2^{(l)}$ and the imaginary part of the spectrum $\sigma_{p,c}$ [$\ln{|\mathrm{Im}(E)|}$] of different eigenstates belonging to the spectrum $\sigma_{p,c}$, respectively, as a function of the real part of the corresponding $E$ and the modulation amplitude $\lambda$ with $\theta=0$, $\epsilon_n^{-}=0$ and $N=200$ under PBCs. The solid lines in Fig. \ref{Fig3} are the analytical solution of mobility edges. The real-complex transition in energy coincides with the localization transition shown in Fig. \ref{Fig3}(b). One can see that our analytical result is in excellent agreement with the numerical results. 
	For the antisymmetric case $\epsilon_n^{+}=0$ obtained by $\phi_b=\pi$ for $\theta=0$, all flat-band states are expelled from their unperturbed energy position $E_0$. Since $\epsilon_n^{-} \ne 0$, at the flat-band energy $E_0$, we can obtain $p_n=f_n=c_n=0$. Therefore, only the trivial state $(p_n,f_n,c_n)=(0,0,0)$ satisfies Eq.(\ref{eq7}) with $\theta=0$ and $\phi_b=\pi$ at the flat-band energy $E_0$. In Figs. \ref{Fig4}(a) and \ref{Fig4}(b), we respectively plot the fractal dimension $D_2^{(l)}$ and $\ln{|\mathrm{Im}(E)|}$ as the function of $\mathrm{Re}(E)$ and $\lambda$ in the case of antisymmetry and $\theta=0$ with $N=200$ under PBCs. In this case, mobility edges can be observed. The system can be described effectively by a non-Hermitian AA chain eigenequation with the nonreciprocal hoppings and the on-site modulation being $\lambda^2(E-K)/(4E) \cos(4\pi\alpha n)$. From Ref. \cite{Jiang2019}, the analytic expression of the mobility edge is
	\begin{equation}
		\lambda_c=\sqrt{\left|\frac{8E_c M}{E_c-K}\right|}. \label{eq11}
	\end{equation}
	The analytic curve of the mobility edge Eq.(\ref{eq11}) is plotted in Fig.\ref{Fig4} marked by the solid lines, displaying agreement with our numerical results.
	
	For the $\theta=\pi$ cases for Eq. (\ref{eq7}), we obtain a tight-binding form containing the $p_n$ variables only:
	\begin{equation}
		\left[\frac{E(E-K)}{2}-1\right]  p_n = \tilde{J}_n e^{-h}p_{n-1} +\tilde{J}_{n+1}e^{h}p_{n+1} +\tilde{\epsilon}_n^{(2)} p_n, \label{eq12}
	\end{equation}
	with
	\begin{equation}
		\tilde{\epsilon}_n^{(2)} =\frac{E-K}{2}\left[\epsilon_n^{+} + \frac{(\epsilon_n^{-})^2}{E-\epsilon_n^{+}-\frac{2}{E-K}}+\frac{(\frac{2}{E-K})^2}{E-\epsilon_{n+1}^{+}-\frac{2}{E-K}}\right],\label{eq13}\\
	\end{equation}
	and 
	\begin{equation}
		\tilde{J}_n=\frac{\epsilon_n^{-}}{E-\epsilon_n^{+}-\frac{2}{E-K}}.\label{eq14}
	\end{equation}
	The reduced topology displays a complex tight-binding form with the on-site modulation term and the hopping terms being $\epsilon_n^{+}$, $\epsilon_n^{-}$, and $K$ dependent, which should present complex localization properties.
	
	When we consider the symmetric case $\epsilon_{n}^{-}=0$ for $\theta=\pi$, the intracell and intercell information can be written as the matrix form:
	\begin{eqnarray}
		\hat{V}^{(1)}=\left(\begin{array}{ccc}0&0&0\\0&0&\sqrt{2}\\0&\sqrt{2}&0\end{array}\right), \quad
		\hat{T}^{(1)}=\left(\begin{array}{ccc}0&0&0\\0&0&0\\ \sqrt{2}&0&0\end{array}\right).\label{eq15}
	\end{eqnarray}
	And the lattice eigenvalue equation similar to Eq. (\ref{eq1}) reads
	\begin{equation}
		E\varphi_n=-\left[\hat{V}^{(1)}\varphi_n+\hat{T}^{(1)}\varphi_{n-1}+\hat{T}^{(1)\dagger}\varphi_{n+1}\right] + \hat{\epsilon}_n^{(1)}\varphi_n, \label{eq16}
	\end{equation}
	with the on-site disorder matrix $\hat{\epsilon}_n^{(1)}=\mathrm{diag}(\epsilon_n^+,\epsilon_n^+,K)$. A new unit cell can be identified considering the connected lattice sites $\tilde{\varphi}_n=(p_{n-1}, f_n, c_n)$, which affirms that the CLS of the disorder-free limit stays in one unit cell. The corresponding information on the intracell and the intercell reads
	\begin{eqnarray}
		\hat{V}^{(2)}=\left(\begin{array}{ccc}0&0&\sqrt{2}\\0&0&\sqrt{2}\\ \sqrt{2}&\sqrt{2}&0\end{array}\right), \quad
		\hat{T}^{(2)}=\left(\begin{array}{ccc}0&0&0\\0&0&0\\0&0&0\end{array}\right), \label{eq17}
	\end{eqnarray}
	with the lattice eigenvalue equation
	\begin{equation}
		E\tilde{\varphi}_n=-\left[\hat{V}^{(2)}\tilde{\varphi}_n+\hat{T}^{(2)}\tilde{\varphi}_{n-1}+\hat{T}^{(2)\dagger}\tilde{\varphi}_{n+1}\right] + \hat{\epsilon}_n^{(2)}\tilde{\varphi}_n,\label{eq18}
	\end{equation}
	where $\hat{\epsilon}_n^{(2)}=\mathrm{diag}(\epsilon_{n-1}^+,\epsilon_n^+,K)$. 
	Observing the geometric information above, one can find that Eq.(\ref{eq18}) displays the vanishing of the hopping between adjacent unit cells and the hopping term only exists within one unit cell. The extensive degeneracy is broken with the energy being modulation-dependent. With the help of the transformation, we display that the symmetric case with $\theta=\pi$ is made of three-site unit cells but with the absence of intercell hoppings, indicating the preservation of the CLSs even in the presence of the disorder. It means that all the states in such case are localized. This conclusion can be also got by setting $\epsilon_n^-=0$ in Eq. (\ref{eq12}), where the hopping term vanishes and it only exists the on-site modulation. Therefore, we can easily conclude that for a finite $\lambda$ in the symmetric case with $\theta=\pi$, all the states are localized with non-degenerate energies.
	
\begin{figure}[!htb]
		\begin{center}
			\includegraphics[width=0.8 \textwidth] {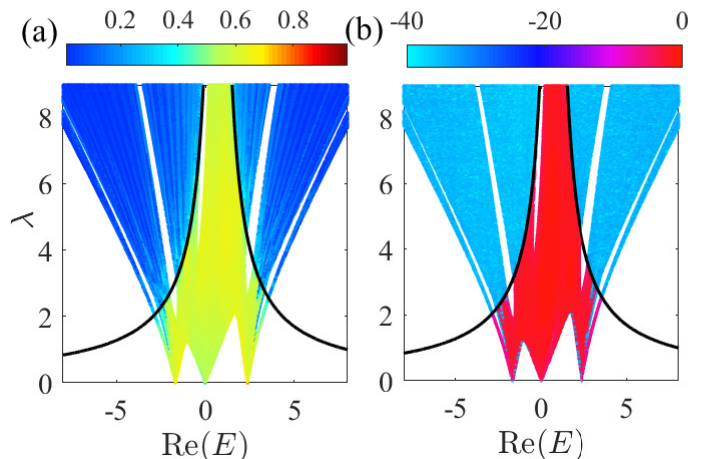}
		\end{center}
		\caption{(Color online) Antisymmetric Case with $\theta=\pi$: (a) The real part of the spectrum as a function of $\lambda$, where the color denotes the value of $D_2^{(l)}$. (b) $\ln{|\mathrm{Im}(E)|}$ as a function of $\lambda$ and $\mathrm{Re}(E)$, where the color denotes the value of  $\ln{|\mathrm{Im}(E)|}$. The black solid lines represent the multifractal-to-localized edges given by Eq. (\ref{eq21}). Here, $N=200$.}\label{Fig5}
	\end{figure}

	For the antisymmetric case $\epsilon_n^{+}=0$ for $\theta=\pi$ , when $E=K$, the tight-binding equation becomes
	\begin{equation}
		-\frac{2K}{\lambda} p_n= \cos(2\pi n\alpha)e^{-h}p_{n-1}+\cos(2\pi (n+1)\alpha)e^h p_{n+1}. \label{eq19}
	\end{equation}
	This model is equivalent to the non-Hermitian off-diagonal Harper model \cite{Tang2021}, where the $E=K$ modes remain multifractal for all the modulation amplitude. For the $E\ne K$ case, the tight-binding equation Eq. (\ref{eq12}) can be given as:
	\begin{eqnarray}
		\tilde{E}p_n &=& \cos(2\pi \alpha n) e^{-h}p_{n-1} \nonumber\\
		&+&\cos[2\pi \alpha (n+1)]e^h p_{n+1}+\frac{\lambda(E-K)}{4}\cos(4\pi \alpha n)p_n, \label{eq20}
	\end{eqnarray}
	with $\tilde{E}=\frac{E^2(E-K)-4E}{2\lambda}-\frac{\lambda(E-K)}{4}$. Eq. (\ref{eq20}) is similar to a generalized Harper model with nonreciprocal hoppings \cite{Tang2021}, but its on-site modulation frequency is twice that of the hoppings. According to Refs. \cite{Dombrowski1978,Marx2017}, it has not been allowed extended states that the mode described by Eq.(\ref{eq20}). We plot the fractal dimension $D_2^{(l)}$ and $\ln{|\mathrm{Im}(E)|}$ as the function of $\mathrm{Re}(E)$ and $\lambda$ in the case of antisymmetry and $\theta=\pi$ with $N=200$ under PBCs, shown in Figs. \ref{Fig5}(a) and \ref{Fig5}(b), respectively. According to the numerical calculation, we can obtain the delocalized-to-localized edges
	\begin{equation}
		\lambda_c = \left|\frac{4M}{E_c-K}\right|,\label{eq21}
	\end{equation}	
	which are plotted in Fig.\ref{Fig5} marked by the solid lines. In the Hermitian limit with $h=0$, Eq.(\ref{eq21}) reduces to $\lambda_c=4/|E_c-K|$, which has been studied numerically in Ref. \cite{Ahmed2022}. One can see that in Fig.~\ref{Fig5}(a), the values of $D_2^{(l)}$ localized in the internal region between the two solid lines are around $0.55$ to $0.7$, indicating that the states in this region may be multifractal under PBCs. 
	
\begin{figure}[!htb]
		\begin{center}
			\includegraphics[width=1 \textwidth] {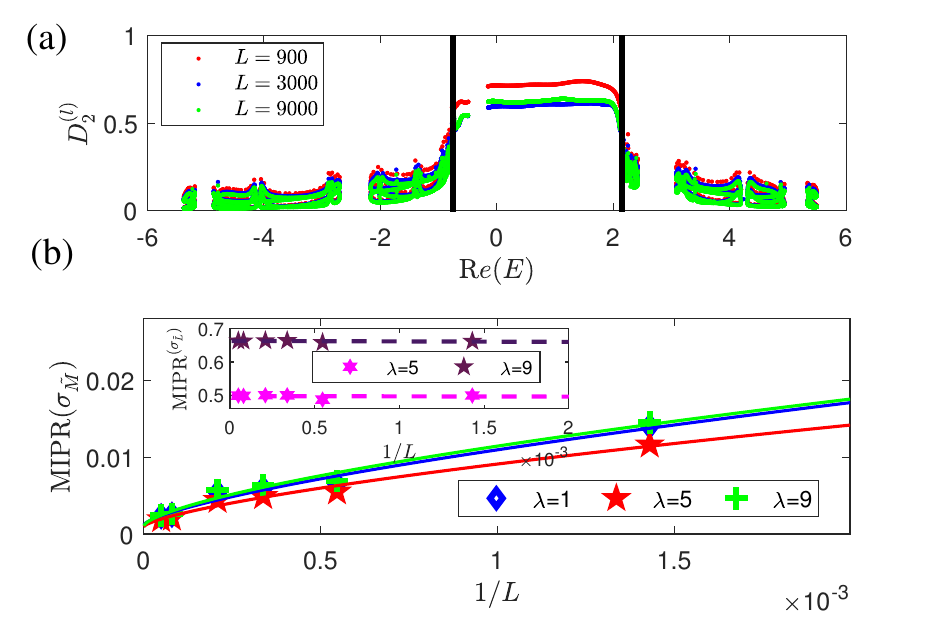}
		\end{center}
		\caption{(Color online) Antisymmetric Case with $\theta=\pi$: (a) $D_2^{(l)}$ for each eigenstate at different system sizes with $\lambda=5$. (b) The scaling of $\mathrm{MIPR}(\sigma_{\tilde{M}})$ for different $\lambda$. The inset shows the scaling of $\mathrm{MIPR}(\sigma_{\tilde{L}})$ for different $\lambda$.}\label{Fig6}
	\end{figure}

	To further determine the localization properties in this case, we consider the fractal dimensions $D_2^{(l)}$ for each eigenstate at different system sizes, which is shown in  Fig. \ref{Fig6}(a) with $\lambda=5$ and $\theta=\pi$ under PBCs. In the finite-size case, the fractal dimensions of the states in the localized regions extrapolate to $0$. In contrast, the fractal dimension's values of the internal region between two analytical edges shown by Eq.(\ref{eq21}) with $\lambda=5$ are away from $0$ and $1$ for different sizes, which implies that the states in this internal region are multifractal, and the analytical edges correspond to the multifractal-to-localized edges. Furthermore, we calculate $\mathrm{MIPR}(\sigma_{\tilde{M}})$ as a function of $1/L$ for different $L$ shown in Fig.\ref{Fig6}(b). When $\lambda=1$, all the states are multifractal, and the corresponding fitting parameters of $\mathrm{MIPR}(\sigma_{\tilde{M}})$ are $\tilde{a}\approx 1.27$, $\tilde{b}\approx 0.705$, and $\tilde{c} \to 0$, respectively. For the $\lambda=5$ case, the mulitfractal-to-localized edges are at $E_{c1}\approx -0.75$ and $E_{c2}\approx 2.15$, and the corresponding fitting function is $\mathrm{MIPR}_{\lambda=5}(\sigma_{\tilde{M}})\approx 1.018 L^{-0.699}$ with $\tilde{c}\to 0$. The mulitfractal-to-localized edges of $\lambda=9$ are at $E_{c1}\approx -0.1$ and $E_{c2} \approx 1.5$. The fitting function is $\mathrm{MIPR}_{\lambda=9}(\sigma_{\tilde{M}})\approx 1.061 L^{-0.670}$ with $\tilde{c}\to 0$. According to the fitting parameters of the mean inverse participation ratios of the internal regions, we can further determine that the states in the internal regions are multifractal. We also show the scaling of $\mathrm{MIPR}(\sigma_{\tilde{L}})$ for $\lambda=5$ and $9$ in the inset of Fig. \ref{Fig6}(b). Both cases display the $M$-independent behavior, and in the thermodynamic limit, $\mathrm{MIPR}(\sigma_{\tilde{L}})$ tend to finite values. Our results imply that the system has multifractal-to-localized edges for the $\theta=\pi$ and antisymmetric case, which can separate the multifractal states from the localized ones.
	
\begin{figure}[!htb]
		\begin{center}
			\includegraphics[width=1 \textwidth] {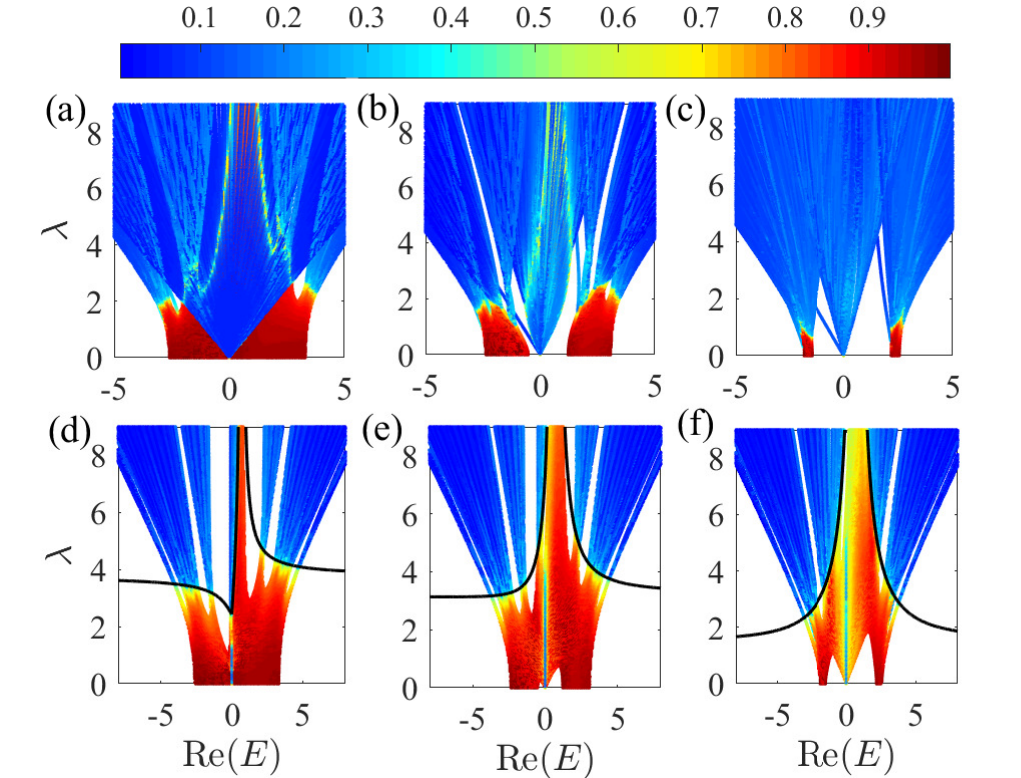}
		\end{center}
		\caption{(Color online) $D_2^{(l)}$ as a function of $\lambda$ and $\mathrm{Re}(E)$ with $N=200$, where the color denotes the value of $D_2^{(l)}$. The top row corresponds to the symmetric cases, and the bottom row corresponds to the antisymmetric cases. From the left column to the right column, $\theta=0.15\pi$, $0.5\pi$, and $0.9\pi$, respectively.}\label{Fig7}
	\end{figure}

	In the following, we consider when $\theta \in (0,\pi)$, the localization properties for both symmetric and antisymmetric cases.	For the symmetric case $\epsilon_n^-=0$, when $\theta=0$, the system presents two independent spectra, $\sigma_f$ and $\sigma_{p,c}$, where the $\sigma_f$ keeps the localized properties and the $\sigma_{p,c}$ displays the mobility edges separating the extended states from the localized ones. Figures \ref{Fig7}(a)-\ref{Fig7}(c) show the fractal dimension $D_2^{(l)}$ of different eigenstates as a function of $\mathrm{Re}(E)$ and the modulation amplitude $\lambda$ for $\epsilon_n^-=0$ with $\theta=0.15\pi$, $0.5\pi$, and $0.9\pi$, respectively. In the small $\theta$ case, the spectra $\sigma_f$ and $\sigma_{p,c}$ begin to couple together and display a weak coupling at the edges of the two spectra. With the increase of $\theta$, in the small $\lambda$ case, the proportion of the extended states gradually decreases in the band-edge regions. However, the band-center region gradually changes from a mixture regime with both extended and localized states to a multifractal regime to a localized regime. When $\theta \to \pi$, all the states become localized for an arbitrary finite $\lambda$. 
	According to our numerical calculation, we can see that, in this case, the system displays a complex localization feature, and the existence of extended, localized, and multifractal regimes is detected. For the antisymmetric case $\epsilon_n^+=0$, we can obtain the exact extended-to-localized edges at $\theta=0$, and when $\theta=\pi$, the multifractal-to-localized edges are given by Eq.(\ref{eq21}). To obtain the localization information for an arbitrary $\theta$, we plot $D_2^{(l)}$ as a function of $\mathrm{Re}(E)$ and the modulation amplitude $\lambda$ for $\epsilon_n^+=0$ with $\theta=0.15\pi$, $0.5\pi$, and $0.9\pi$ shown in Figs. \ref{Fig7}(d)-\ref{Fig7}(f), respectively. We find that the delocalization-to-localization transition can be described by the equation
	\begin{equation}
		\lambda_c=2\sqrt{\Big|\frac{E_c M}{E_c-K}\Big|\sqrt{2(1+\cos\theta)}+\frac{2(1-\cos\theta)M^2}{(E_c-K)^2}},\label{eq22}
	\end{equation}
	which is plotted as the black solid lines in Figs. \ref{Fig7}(d)-\ref{Fig7}(f). Despite the lack of analytical proof, the relation Eq.(\ref{eq22}) works well in separating the delocalized and localized regimes for different $\theta$. Eq.(\ref{eq22}) can be considered an empirical combination of the corresponding analytical results under different limitations. When $\theta \to 0$, Eq.(\ref{eq22}) reduces to Eq.(\ref{eq11}), and for $\theta \to \pi$, Eq.(\ref{eq22}) reduces to Eq.(\ref{eq21}). Our results also suit the Hermitian cases with $h=0$. Moreover, as seen in Figs. \ref{Fig7}(d)-\ref{Fig7}(e), the multifractal states are induced by increasing $\theta$, and for an intermediate $\theta$, it displays an extended and multifractal mixture in the band-center region.
	
	\begin{figure}[!htb]
		\begin{center}
			\includegraphics[width=0.8 \textwidth] {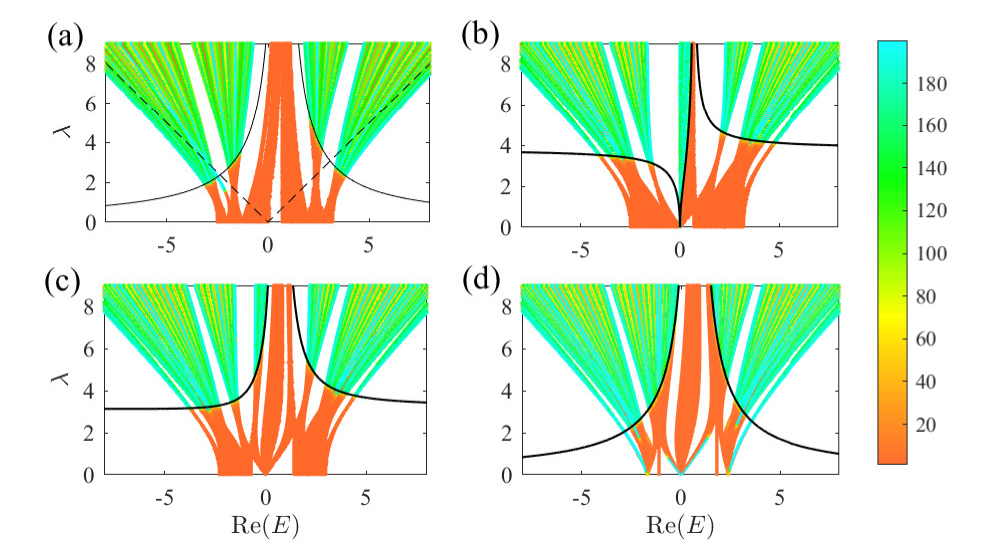}
		\end{center}
		\caption{(Color online) MCMs as functions of $\lambda$ and $\mathrm{Re}(E)$ with $N=200$, $h=0.6$, and $K=0.7$ under OBCs are shown for: (a) the symmetric case with $\theta=0$, (b) the antisymmetric case with $\theta=0$, (c) the antisymmetric case with $\theta=0.5\pi$, and (d) the antisymmetric case with $\theta=\pi$, respectively. The black solid lines correspond to the analytical results for mobility edges. \label{Fig8}}
	\end{figure} 
	
	The entire paper focuses on the PBC cases, and we find that our analytical results align well with numerical calculations using IPR and fractal dimension. However, due to the presence of non-reciprocal coupling, the system may exhibit the skin effect under OBCs, leading to localized features. This raises an important question: Do the localization properties of our system change under OBCs? Traditional order parameters like the IPR and fractal dimension struggle to distinguish between states exhibiting the skin effect and those that are conventionally localized.
	
	To address this, we introduce the mean center of mass (MCM), defined as $\mathrm{MCM}=\sum_{n=1}^{N}n(|\psi_n^{a}|^2+|\psi_n^{b}|^2+|\psi_n^{c}|^2)$ \cite{Li2024} under OBCs to differentiate skin effect states from normally localized ones. Figure (\ref{Fig8}) shows MCMs as functions of $\lambda$ and $\mathrm{Re}(E)$ for different values of $\theta$ and the symmetries of the extended modulations with $N=200$, $h=0.6$, and $K=0.7$ under OBCs. The black solid lines represent the analytical results for mobility edges discussed earlier. The minimum value of MCMs in the central region of the analytical solution indicates that the wave function is localized at the left boundary due to nonreciprocal hopping, exhibiting characteristics of the skin effect. In contrast, the outer region shows much larger values, indicating localized features that are unaffected by non-reciprocal hopping. This demonstrates that the delocalized states under PBCs transform into non-Hermitian skin states under OBCs, while the localized states retain their localization features under OBCs.

	\section{Experimental proposal}
	
	\begin{figure}[!htb]
		\begin{center}
			\includegraphics[width=0.8 \textwidth] {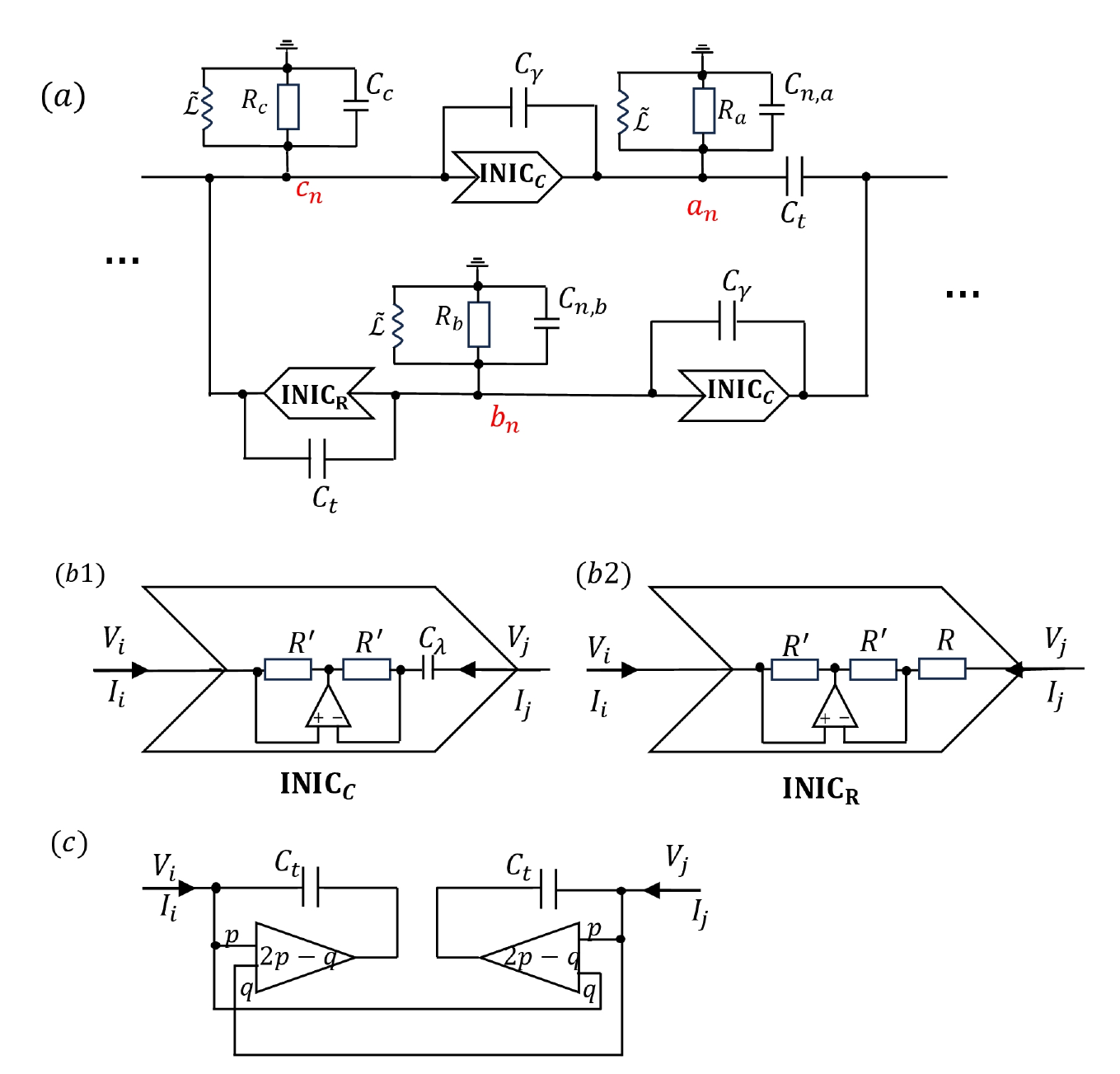}
		\end{center}
		\caption{(Color online) (a) Electrical circuit implementation of the model in Eq. (\ref{eq1}). Details of $\mathrm{INIC}_R$ (b1) and $\mathrm{INIC}_R$ (b2). (c) The equivalent negative impedance between two free terminals, where the markings on the ideal amplifier represent the relationship between output voltage and input voltage.\label{Fig9}}
	\end{figure}
	
	The localization-delocalization transitions induced by quasiperiodic modulations in a non-Hermitian flat-band diamond lattice can be experimentally observed in electrical circuits. We designed a non-Hermitian electrical circuit in Fig. \ref{Fig9}, corresponding to the model in Eq. (\ref{eq1}). In Fig. \ref{Fig9}, the nonreciprocal hopping and the hopping with a magnetic flux $\theta$ terms can be realized with the aid of negative impedance converters through current inversion (INIC)\cite{Li2024}, as shown in Figs. \ref{Fig9}(b1) ($\mathrm{INIC}_C$) and \ref{Fig9}(b2) ($\mathrm{INIC}_R$), respectively. The nonreciprocal hopping between $c_n$ and $a_n$, and between $b_n$ and $c_{n+1}$, are simulated by combining a normal capacitor $C_{\gamma}$ and an $\mathrm{INIC}_C$. The $\mathrm{INIC}_C$ consists of one capacitor $C_{\lambda}$, one operational amplifier, and two resistors with equal resistance values $R^{\prime}$. When the current flows from left to right, the capacitance between $c_n$ and $a_n$, and between $b_{n}$ and $c_{n+1}$, is $C_{\gamma}-C_{\lambda}$. If the current runs in the opposite direction, the capacitance will be $C_{\gamma}+C_{\lambda}$. When $\theta\in [0,\pi/2)$, the hopping with a magnetic flux $\theta$ between $c_n$ and $b_n$ is simulated by combining a normal capacitor $C_{t}$ and an $\mathrm{INIC}_R$. And for $\theta\in (\pi/2,\pi]$, we replace the normal capacitor $C_{t}$ with a two-terminal configuration to realize achieve negative capacitor $-C_t$ shown in Fig. \ref{Fig9}(c), which consists of two capacitors and two operational amplifiers. The $\mathrm{INIC}_R$ consists of one resistor $R$, one operational amplifier, and two resistors with equal resistance values $R^{\prime}$. When the current runs from left to right between $c_n$ and $b_n$, the value of the effective admittance is $i\omega Z_{\theta}$, where $Z_{\theta}=C_t+\frac{1}{i\omega R}$ 
	for $\theta\in [0,\pi/2)$ and $Z_{\theta}=-C_t+\frac{1}{i\omega R}$ 
	for $\theta\in (\pi/2,\pi]$. When the current runs in the opposite direction, the value of the effective admittance is $i\omega Z_{\theta}^{\dagger}$. 
	 The on-site potentials at each site is simulated by grounding each node with three suitable devices chosen according to the values of their impedance. This model described by Eq. (\ref{eq1}) can be represented by the circuit Laplacian $\mathcal{J}(\omega)$ of the circuit. The Laplacian is defined as the response of the grounded-voltage vector $V$ to the vector $I$ of input current by
	\begin{equation}
		I(\omega)=\mathcal{J}(\omega)V(\omega).\label{eq23}
	\end{equation}
	Using Eq. (\ref{eq23}), the current of each node within the unit cell can be expressed as
	\begin{small}
		\begin{eqnarray}
			\dot{I}_{n,a}=&i\omega(C_{\gamma}-C_{\lambda})V_{n,c}+i\omega C_tV_{n+1,c}\nonumber\\
			&-\left[i\omega (C_{\gamma}-C_{\lambda}+C_t+C_{n,a})+\frac{1}{R_{a}}+\frac{1}{i\omega \tilde{\mathcal{L}}}\right]V_{n,a},\nonumber\\
			\dot{I}_{n,b}=&i\omega(C_{\gamma}+C_{\lambda})V_{n+1,c}+i\omega Ze^{-i\theta}V_{n,c}\nonumber\\
			&-\left[i\omega (C_{\gamma}+C_{\lambda}+C_t+C_{n,b}) +\frac{1}{R_{b}}+\frac{1}{R}+\frac{1}{i\omega \tilde{\mathcal{L}}}\right]V_{n,b},\nonumber\\
			\dot{I}_{n,c}=&i\omega(C_{\gamma}+C_{\lambda})V_{n,a}+i\omega(C_{\gamma}-C_{\lambda})V_{n-1,b}+i\omega Ze^{i\theta}V_{n,b}\nonumber\\
			&+i\omega C_tV_{n-1,a}-\left[i\omega (2C_{\gamma}+2C_t+C_{c})+\frac{1}{R_{c}}+\frac{1}{R}+\frac{1}{i\omega \tilde{\mathcal{L}}}\right]V_{n,c},\label{eq24}
		\end{eqnarray}
	\end{small}
	where the vector $I_{n,\beta}$ and $V_{n,\beta}$ represent the node currents and voltages of the $\beta$th sublattices within the $n$th unit cell with $\beta=\{a,b,c\})$, respectively. Consequently, the targeted circuit Laplacian $\mathcal{J}(\omega)$ is employed to simulate the model in Eq. (\ref{eq24}), as follows:
	\begin{small}
		\begin{eqnarray}
			\mathcal{J}(\omega)=&i\omega \left(\begin{array}{ccccccc} C_{\Delta a,1} & 0 & C_{\gamma}-C_{\lambda} & \dots & 0 & 0 & 0 \\
				0 & C_{\Delta b,1} & Z_{\theta} & \dots & 0 & 0 & 0 \\
				C_{\gamma}+C_{\lambda} & Z_{\theta}^{\dagger} & C_{\Delta c} & \dots & C_t & C_{\gamma}-C_{\lambda} & 0 \\
				\vdots & \vdots & \vdots & \ddots & \vdots & \vdots & \vdots \\ 
				0 & 0 & C_t & \dots & C_{\Delta a,n} & 0 & C_{\gamma}-C_{\lambda}\\
				0 & 0 & C_{\gamma}+C_{\lambda} & \dots & 0 & C_{\Delta b,n} & Z_{\theta}\\
				0 & 0 & 0 & \dots & C_{\gamma}+C_{\lambda} & Z_{\theta}^{\dagger}& C_{\Delta c}\end{array}\right)\nonumber\\
			&-i\omega\Big[C_{\gamma}+C_t+\frac{1}{i\omega R_{\Delta}}-\frac{1}{\omega^2 \tilde{\mathcal{L}}}\Big]\mathcal{I},
			\label{eq25}
		\end{eqnarray}
	\end{small}
	where $\mathcal{I}$ represents an identity matrix,  $C_{\Delta a,n}=-C_{n,a}+C_{\lambda}$,  $C_{\Delta b,n}=-C_{n,b}-C_{\lambda}$, and  $C_{\Delta c}=-C_{\gamma}-C_{t}-C_{c}$. Furthermore, the introduction of $\frac{1}{R_{a}}=\frac{1}{R}+\frac{1}{R_{b}}=\frac{1}{R_{c}}+\frac{1}{R}=\frac{1}{R_{\Delta}}$ and inductance $\tilde{\mathcal{L}}$ shifts the admittance spectrum as a whole in the complex plane, ensuring that the circuit's response does not diverge.  Hence the Hamiltonian for the non-Hermitian flat-band system with quasiperiodic modulations is achieved. The energy spectrum of the system can be obtained from the admittance spectrum of the circuit, and the distribution of states can be detected by measuring the voltage at each node.

	\section{Conclusion} 
	This paper investigates the effects of quasiperiodic AA modulations on a one-dimensional non-Hermitian diamond lattice featuring flat bands. For $\theta=0$, the system exhibits a single flat band in the crystalline limit. Symmetric and antisymmetric modulations lead to the emergence of exact mobility edges. However, when $\theta=\pi$, the system becomes an ABF system without disorder. In this case, symmetric disorder perturbs the degeneracy completely, while the corresponding CLSs persist. In contrast, the application of antisymmetric modulation disrupts compact localization, giving rise to multifractal states for any finite modulation amplitude. We employ numerical calculations to derive the expression for the transition from multifractal to localized states, offering insights into this multifractal-to-localized edge. Furthermore, we explore cases where $\theta$ lies within the range $(0, \pi)$, revealing complex localization features within the system. Finally, we design a classic electrical circuit to realize our quasiperiodic modulated non-Hermitian flat-band system.
	
	This work focuses on the intricate interplay between quasiperiodic modulations and nonreciprocal hopping in a flat-band system, an area that has been largely unexplored. By addressing how different symmetries of introduced modulations and synthetic magnetic flux parameters influence localization, our study contributes to a deeper understanding of the fundamental properties of non-Hermitian quasiperiodic systems. Moreover, our proposed experimental realization in electrical circuits provides a practical method for exploring these phenomena, thereby broadening the potential for future research in non-Hermitian physics.
	
	\bigskip
	Z. Xu is supported by the NSFC (Grant No. 12375016), Fundamental Research Program of Shanxi Province (Grant No. 20210302123442), and the Beijing National Laboratory for Condensed Matter Physics (Grant No. 2023BNLCMPKF001). X. Xia is supported by the NSFC (Grant No. 12301218). This work is also supported by NSF for Shanxi Province Grant No. 1331KSC.
	\clearpage
	
	\appendix
	
	\section{Derivation of CLSs}
	\setcounter{section}{1}
	\setcounter{figure}{0} \setcounter{table}{0} %
	\renewcommand{\thefigure}{A\arabic{figure}}
	\renewcommand{\thetable}{A\Roman{table}}
	In this Appendix, we present a detailed derivation of the CLSs in Fig. \ref{Fig2}(c)-Fig. \ref{Fig2}(e), following the method proposed in Ref. \cite{Maimaiti2021}. We can search for the non-Hermitian Hamiltonian Eq. (\ref{eq1}) with the CLS $\Psi_{\mathrm{CLS}}=(\psi_1,\psi_2,\dots,\psi_u)$ of size $u$ (occupying exactly unit cells). The CLS $\Psi_{\mathrm{CLS}}$ is an eigenvector of the $u\times u$ block tridiagonal matrix
  \begin{eqnarray}
	\mathcal{H}_u=\left(\begin{array}{cccc}
	\hat{V} & \hat{T}_2 & 0 & 0\\
	\hat{T}_1 & \hat{V} & \cdots & 0\\
	\vdots & \ddots & \ddots & \vdots \\
	0 & \cdots & \hat{T}_1 & \hat{V}\\
    \end{array}\right),
	\label{A.1}
  \end{eqnarray}
  with eigenenergy $E_{\mathrm{FB}}$ and
  \begin{eqnarray}
  	\hat{V}=\left(\begin{array}{ccc}0&0&e^{-h}\\0&0&e^{-i\theta}\\e^h&e^{i\theta}&0\end{array}
  	\right),\hat{T}_1=\left(\begin{array}{ccc}0&0&0\\0&0&0\\1&e^{-h}&0\\ \end{array}
  	\right),
  	\hat{T}_2=\left(\begin{array}{ccc}0&0&1\\0&0&e^h\\0&0&0\end{array}
  	\right),
  	\label{A.2}
  \end{eqnarray}
  Out of the $u$ eigenevctors of the Hamiltonian, the CLS is selected by the destructive interference conditions
  \begin{eqnarray}
  	\hat{T}_2 |\psi_1\rangle =\hat{T}_1 |\psi_u\rangle =0 \label{A.3}
  \end{eqnarray}
  which ensure that the eigenstate remains compactly localized under the action of the Hamiltonian $\mathcal{H}_u$. One can solve the eigenproblem for the CLS as follows:
  \begin{eqnarray}
  	\hat{T}_2|\psi_2\rangle = (E_{\mathrm{FB}}-\hat{V})|\psi_1\rangle, \nonumber \\
  	\hat{T}_1|\psi_{n-1}\rangle+\hat{T}_2|\psi_{n+1}\rangle = (E_{\mathrm{FB}}-\hat{V})|\psi_{n}\rangle, \quad 2\le n \le u-1 \nonumber \\
  	\hat{T}_1|\psi_{u-1}\rangle = (E_{\mathrm{FB}}-\hat{V})|\psi_u\rangle, \nonumber \\
  	\hat{T}_2 |\psi_1\rangle = \hat{T}_1 |\psi_u\rangle =0, \nonumber\\
  	|\psi_n\rangle = 0, \quad n<0, n>u. \label{A.4}
  \end{eqnarray}
  One can obtain the expression of the CLSs that satisfy Eq. (\ref{A.4}).
  
  According to our numerical results, we find that in our cases, the CLS occupies $u=2$ unit cells at $E_0=0$ for $\theta \neq 0$ and $E_{\pm}=\pm 2J$ for $\theta = \pi$, and the CLS at $E_0=0$ for $\theta =0$ occupies $u=1$ unit cell. For the $\theta = \pi$ case with $u=2$, the block tridiagonal matrix $\mathcal{H}_u$ reads:
	  \begin{eqnarray}
		\mathcal{H}_u=\left(\begin{array}{cc}
			\hat{V} & \hat{T}_2 \\
			\hat{T}_1 & \hat{V} \\
		\end{array}\right),
		\label{A.5}
	\end{eqnarray}
	with eigenenergy $E_{FB}=E_0$ and $E_{\pm}$, and 
	\begin{eqnarray}
		\hat{V}=\left(\begin{array}{ccc}0&0&e^{-h}\\0&0&-1\\e^h&-1&0\end{array}
		\right),\hat{T}_1=\left(\begin{array}{ccc}0&0&0\\0&0&0\\1&e^{-h}&0\\ \end{array}
		\right),
		\hat{T}_2=\left(\begin{array}{ccc}0&0&1\\0&0&e^h\\0&0&0\end{array}
		\right).
		\label{A.6}
	\end{eqnarray}
	The eigenproblem for the CLS Eq. (\ref{A.5}) reads:
		\begin{eqnarray}
		\hat{T}_2 |\psi_2\rangle =(E_{FB}-\hat{V}) |\psi_1\rangle ,\nonumber\\
		\hat{T}_1 |\psi_1\rangle =(E_{FB}-\hat{V}) |\psi_2\rangle ,\nonumber\\
		\hat{T}_2 |\psi_1\rangle =0, \quad \hat{T}_1 |\psi_2\rangle =0. 
		\label{A.7}
	\end{eqnarray}
    When $E_{FB}=E_0$, we can obtain $|\psi_1\rangle={(1,e^h,0)}^T,\quad |\psi_2\rangle={(-e^{-h},1,0)}^T$, and for $E_{FB}=E_{\pm}$, $|\psi_1\rangle={(1,e^h,0)}^T,\quad |\psi_2\rangle={(e^{-h},-1,\pm 2)}^T$. Hence, the CLS of $\theta=\pi$ at $E_0$ is $\Psi_{\mathrm{CLS}}=\{1,e^h,0,-e^{-h},1,0\}$, and at $E_{\pm}$ is $\Psi_{\mathrm{CLS}}=\{1,e^h,0,e^{-h},-1,\pm 2\}$. Following similar processes, one can obtain the CLS of $\theta \neq 0$ at $E_{0}$ is $\Psi_{\mathrm{CLS}}=\{e^{i\theta},-e^h,0,e^{-h},-1,0\}$. For the $\theta = 0$ and $E_{\mathrm{FB}}=E_0$ case with $u=1$, the eigenvector of $\hat{V}$ is the CLS $\Psi_{\mathrm{CLS}}=(\psi_1)^T=\{1,-e^h,0\}$.
   
    \begin{figure}[!htb]
   	\begin{center}
   		\includegraphics[width=0.8 \textwidth] {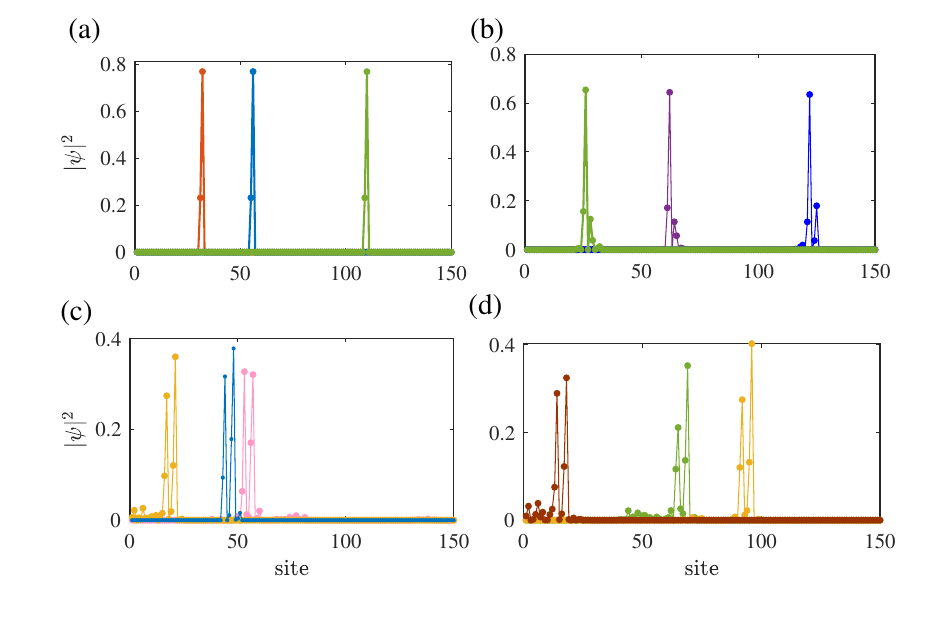}
   	\end{center}
   \caption{(Color online) Density distributions of three randomly selected eigenstates $|\psi|^2$ under OBCs for (a) $\theta = 0, E_0=0$ , (b) $\theta = \pi, E_0=0$, (c) $\theta = \pi, E_{+}=2J$, and (d) $\theta = \pi, E_{-}=-2J$.\label{FigA1}}
   \end{figure} 
   
   Due to the macroscopic degeneracy in flat-band systems, arbitrary linear combinations of the CLSs for each flat band are the solutions to the system. To mitigate the effects of this degeneracy, we introduce a tiny random disorder at each site in our numerical calculations under OBCs for the states at $E_{\mathrm{FB}}$. Figure \ref{FigA1} shows the density distributions $|\psi|^2$ for three randomly selected eigenstates in different flat bands with $h=0.6$, $N=50$, and an amplitude of the uncorrelated random disorder set to $10^{-12}$ under OBCs. As shown in Fig. \ref{FigA1}, the density distributions in these cases are in excellent agreement with our analytical results presented in the main text. This suggests that the density distributions of CLSs retain their characteristic features under OBCs, even in the presence of the non-reciprocal coupling terms. In contrast, states in dispersive bands hosting the non-reciprocal coupling terms exhibit distinct behaviors depending on the boundary conditions.
   
  \section{Derivation of the rotation matrix $\hat{U}$}
   \setcounter{section}{2}
   \setcounter{figure}{0} \setcounter{table}{0} %
   \renewcommand{\thefigure}{B\arabic{figure}}
   \renewcommand{\thetable}{B\Roman{table}}
   In this Appendix, we provide the derivation of the rotation matrix $\hat{U}$ from Eq. (\ref{eq4}) in the main text, following the method proposed in Ref. \cite{Danieli2021}. The core idea of this method is that by applying a sequence of local transformations, one can transform a Hamiltonian with flat bands into a set of decoupled sites. This implies that all CLSs in these systems are orthogonal and can be represented with nonzero amplitude in a single unit cell.
   
   Consider the case where $\theta=\pi$ first. We define $\nu \times u$ CLS tensors $A_{\nu}$ that parametrize the CLSs with $\nu$ flat bands. For the $\theta=\pi$ case with $u=2$, according to the calculations in Appendix A, the three tensors corresponding to the three flat bands are given by: \cite{Vidal2000,Danieli2021}:
   	\begin{eqnarray}
   	A_{1}=\left(\begin{array}{ccc}1&-e^{-h}\\e^{h}&1\\0&0\end{array}
   	\right),
   	A_{2}=\left(\begin{array}{ccc}1&e^{-h}\\e^{h}&-1\\0&2 \end{array}\right),
   	A_{3}=\left(\begin{array}{ccc}1&e^{-h}\\e^{h}&-1\\0&-2\end{array}\right).
   	\label{B1}
   \end{eqnarray}
   Based on the formation of $A_{\nu}$, we set the real transformation $\hat{U}$ as follows:
    \begin{eqnarray}
    	\left(\begin{array}{c}\varphi_{n,a} \\ \varphi_{n,b} \\ \varphi_{n,c} \end{array} \right)= \hat{U}  \left( \begin{array}{c} \psi_{n,a} \\ \psi_{n,b} \\ \psi_{n,c} \end{array} \right), \quad \hat{U}=\left(\begin{array}{ccc}p&q&0\\w&v&0\\0&0&s\end{array}
    	\right).\label{B2}
    \end{eqnarray}
   Applying the transformation $\hat{U}$, the CLS tensors become:
   \begin{eqnarray}
   	B_{1}=\hat{U}A_1=\left(\begin{array}{ccc}p+e^{h}q&-e^{-h}p+q\\w+e^{h}v&-e^{-h}w+v\\0&0\end{array}
   	\right),\end{eqnarray}\label{B3}
   \begin{eqnarray}
   	B_{2}=\hat{U}A_2=\left(\begin{array}{ccc}p+e^{h}q&e^{-h}p-q\\w+e^{h}v&e^{-h}w-v \\ 0&2s \end{array}\right),\label{B4}
   \end{eqnarray}
   \begin{eqnarray}
   	B_{3}=\hat{U}A_3=\left(\begin{array}{ccc}p+e^{h}q&e^{-h}p-q\\w+e^{h}v&e^{-h}w-v \\ 0&-2s \end{array}\right).
   	\label{B5}
   \end{eqnarray}
   To achieve a unit-cell redefinition, the new tensors $B_{\nu}$ must satisfy:
   	\begin{eqnarray}
   	\left(p+e^{h}q\right)\left(e^{-h}p-q\right)=0,\quad\left(p+e^{h}q\right)\pm\left(e^{-h}p-q\right)\neq 0,\nonumber\\
   	\left(w+e^{h}v\right)\left(e^{-h}w-v\right)=0,\quad\left(w+e^{h}v\right)\pm\left(e^{-h}w-v\right)\neq 0.\label{B6}
   \end{eqnarray}
   Combining with the unitary condition, 
    \begin{eqnarray}
   	\hat{U}^2=\left(\begin{array}{ccc}p^2+qw&pq+qv&0\\pw+vw&pw+v^2&0\\0&0&s^2\end{array}\right)= \left(\begin{array}{ccc}1&0&0\\0&1&0\\0&0&1\end{array}\right), \label{B7}
   \end{eqnarray}
    we find 
    \begin{eqnarray}
    	\hat{U}=\frac{1}{\sqrt{2}}\left(\begin{array}{ccc}1&e^{-h}&0\\e^{h}&-1&0\\0&0&\sqrt{2}\end{array}\right), \label{B8}
   \end{eqnarray}
   and
   	\begin{eqnarray}
   	\tilde{B}_{1}=\sqrt{2}\left(\begin{array}{ccc}1&0\\0&-1\\0&0\end{array}
   	\right),
   	\tilde{B}_{2}=\sqrt{2}\left(\begin{array}{ccc}1&0\\0&1\\0&\sqrt{2} \end{array}\right),
   	\tilde{B}_{3}=\sqrt{2}\left(\begin{array}{ccc}1&0\\0&1\\0&-\sqrt{2}\end{array}\right).
   	\label{B9}
   \end{eqnarray}
 All the new tensors $B_{\nu}$ share the same pattern for the zero elements, which allows us to redefine the unit cell as $\{\phi_{n-1,a},\phi_{n,b},\phi_{n,c}\}^T$ so that the CLSs fit into a single unit cell after the redefinition. The CLS after the unit-cell redefinition are
 \begin{eqnarray}
 	C_{1}=\sqrt{2}\left(\begin{array}{ccc}1&0\\-1&0\\0&0\end{array}
 	\right),
 	C_{2}=\sqrt{2}\left(\begin{array}{ccc}1&0\\1&0\\ \sqrt{2}&0 \end{array}\right),
 	C_{3}=\sqrt{2}\left(\begin{array}{ccc}1&0\\1&0\\-\sqrt{2}&0\end{array}\right).
 	\label{B10}
 \end{eqnarray}
  The new CLS $C_{\nu}$ effectively occupy a single unit cell, which directly transforms the system into decoupled sites. Thus, this transformation reduces the original diamond chain Hamiltonian to an ABF system with CLSs of size $u = 1$. It is straightforward to verify that the transformation matrix $\hat{U}$ can reduce the CLS with $E_0$ at $\theta=0$ to decoupled sites from dispersive ones. However, for arbitrary $\theta\ne 0,\pi$, this matrix $\hat{U}$ fails to transform the CLS with $E_0$ into decoupled sites, indicating that $\hat{U}$ is invalid for these cases. Fortunately, we have derived a rational expression for the delocalization-to-localization transition Eq. (\ref{eq22}), which shows excellent agreement with our numerical results.

\end{document}